\magnification 1200

%
%
\newdimen\FigSize       \FigSize=.9\hsize 
%
\newskip\abovefigskip   \newskip\belowfigskip
\gdef\epsfig#1;#2;{\par\vskip\abovefigskip\penalty -500
   {\everypar={}\epsfxsize=#1\nd \centerline{\epsfbox{#2}}}%
    \vskip\belowfigskip}%
%
\newskip\figtitleskip
\gdef\tepsfig#1;#2;#3{\par\vskip\abovefigskip\penalty -500
   {\everypar={}\epsfxsize=#1\nd
    \vbox
      {\centerline{\epsfbox{#2}}\vskip\figtitleskip
       \centerline{\figtitlefont#3}}}%
    \vskip\belowfigskip}%
%
\newcount\FigNr \global\FigNr=0
\gdef\nepsfig#1;#2;#3{\global\advance\FigNr by 1
   \tepsfig#1;#2;{Figure\space\the\FigNr.\space#3}}%
%
%
%
\gdef\ipsfig#1;#2;{
   \midinsert{\everypar={}\epsfxsize=#1\nd
              \centerline{\epsfbox{#2}}}
   \endinsert}%
%
\gdef\tipsfig#1;#2;#3{\midinsert
   {\everypar={}\epsfxsize=#1\nd
    \vbox{\centerline{\epsfbox{#2}}%
          \vskip\figtitleskip
          \centerline{\figtitlefont#3}}}\endinsert}%
%
\gdef\nipsfig#1;#2;#3{\global\advance\FigNr by1%
  \tipsfig#1;#2;{Figure\space\the\FigNr.\space#3}}%
\newread\epsffilein    
\newif\ifepsffileok    
\newif\ifepsfbbfound   
\newif\ifepsfverbose   
\newdimen\epsfxsize    
\newdimen\epsfysize    
\newdimen\epsftsize    
\newdimen\epsfrsize    
\newdimen\epsftmp      
\newdimen\pspoints     
\pspoints=1bp          
\epsfxsize=0pt         
\epsfysize=0pt         
\def\epsfbox#1{\global\def\epsfllx{72}\global\def\epsflly{72}%
   \global\def\epsfurx{540}\global\def\epsfury{720}%
   \def\lbracket{[}\def\testit{#1}\ifx\testit\lbracket
   \let\next=\epsfgetlitbb\else\let\next=\epsfnormal\fi\next{#1}}%
\def\epsfgetlitbb#1#2 #3 #4 #5]#6{\epsfgrab #2 #3 #4 #5 .\\%
   \epsfsetgraph{#6}}%
\def\epsfnormal#1{\epsfgetbb{#1}\epsfsetgraph{#1}}%
\def\epsfgetbb#1{%
%
%
\openin\epsffilein=#1
\ifeof\epsffilein\errmessage{I couldn't open #1, will ignore it}\else
%
%
   {\epsffileoktrue \chardef\other=12
    \def\do##1{\catcode`##1=\other}\dospecials \catcode`\ =10
    \loop
       \read\epsffilein to \epsffileline
       \ifeof\epsffilein\epsffileokfalse\else
%
%
          \expandafter\epsfaux\epsffileline:. \\%
       \fi
   \ifepsffileok\repeat
   \ifepsfbbfound\else
    \ifepsfverbose\message{No bounding box comment in #1; using
defaults}\fi\fi
   }\closein\epsffilein\fi}%
%
%
\def\epsfsetgraph#1{%
   \epsfrsize=\epsfury\pspoints
   \advance\epsfrsize by-\epsflly\pspoints
   \epsftsize=\epsfurx\pspoints
   \advance\epsftsize by-\epsfllx\pspoints
%
%
   \epsfxsize\epsfsize\epsftsize\epsfrsize
   \ifnum\epsfxsize=0 \ifnum\epsfysize=0
      \epsfxsize=\epsftsize \epsfysize=\epsfrsize
%
arithmetic!
%
     \else\epsftmp=\epsftsize \divide\epsftmp\epsfrsize
       \epsfxsize=\epsfysize \multiply\epsfxsize\epsftmp
       \multiply\epsftmp\epsfrsize \advance\epsftsize-\epsftmp
       \epsftmp=\epsfysize
       \loop \advance\epsftsize\epsftsize \divide\epsftmp 2
       \ifnum\epsftmp>0
          \ifnum\epsftsize<\epsfrsize\else
             \advance\epsftsize-\epsfrsize \advance\epsfxsize\epsftmp
\fi
       \repeat
     \fi
   \else\epsftmp=\epsfrsize \divide\epsftmp\epsftsize
     \epsfysize=\epsfxsize \multiply\epsfysize\epsftmp
     \multiply\epsftmp\epsftsize \advance\epsfrsize-\epsftmp
     \epsftmp=\epsfxsize
     \loop \advance\epsfrsize\epsfrsize \divide\epsftmp 2
     \ifnum\epsftmp>0
        \ifnum\epsfrsize<\epsftsize\else
           \advance\epsfrsize-\epsftsize \advance\epsfysize\epsftmp \fi
     \repeat
   \fi
%
%
   \ifepsfverbose\message{#1: width=\the\epsfxsize,
height=\the\epsfysize}\fi
   \epsftmp=10\epsfxsize \divide\epsftmp\pspoints
   \vbox to\epsfysize{\vfil\hbox to\epsfxsize{%
      \includegraphics{#1}%
      \hfil}}%
\epsfxsize=0pt\epsfysize=0pt}%
%
%
{\catcode`\%=12
\global\let\epsfpercent=
%
%
\long\def\epsfaux#1#2:#3\\{\ifx#1\epsfpercent
   \def\testit{#2}\ifx\testit\epsfbblit
      \epsfgrab #3 . . . \\%
      \epsffileokfalse
      \global\epsfbbfoundtrue
   \fi\else\ifx#1\par\else\epsffileokfalse\fi\fi}%
%
%
\def\epsfgrab #1 #2 #3 #4 #5\\{%
   \global\def\epsfllx{#1}\ifx\epsfllx\empty
      \epsfgrab #2 #3 #4 #5 .\\\else
   \global\def\epsflly{#2}%
   \global\def\epsfurx{#3}\global\def\epsfury{#4}\fi}%
%
%
\def\epsfsize#1#2{\epsfxsize}%
%
%

\epsfverbosetrue                        
\abovefigskip=\baselineskip             
\belowfigskip=\baselineskip             
\global\let\figtitlefont\bf             
\global\figtitleskip=.5\baselineskip    

\font\tenmsb=msbm10   
\font\sevenmsb=msbm7
\font\fivemsb=msbm5
\newfam\msbfam
\textfont\msbfam=\tenmsb
\scriptfont\msbfam=\sevenmsb
\scriptscriptfont\msbfam=\fivemsb
\def\Bbb#1{\fam\msbfam\relax#1}
\let\nd\noindent 

\def\natural{{\rm I\kern-.18em N}}

\def\s{{\cal S}}
\def\eps{{\epsilon}}

\def\R{{\Bbb R}}
\def\S{{S}}
\def\s{{s}}

\def\chix{{\raise.5ex\hbox{$\chi$}}}
\def\chixa{{\chix\lower.2em\hbox{$_A$}}}

\def\real{{\rm I\kern-.2em R}}
\def\integer{{\rm Z\kern-.32em Z}}
\def\complex{\kern.1em{\raise.47ex\hbox{
            $\scriptscriptstyle |$}}\kern-.40em{\rm C}}
\def\vs#1 {\vskip#1truein}
\def\hs#1 {\hskip#1truein}

\def\Month{\ifcase\number\month \relax\or January \or February \or
  March \or April \or May \or June \or July \or August \or September
  \or October \or November \or December \else \relax\fi }
\def\date{\Month \the\day, \the\year}

  \hsize=6truein        \hoffset=.25truein 
  \vsize=8.8truein      
  \pageno=1     \baselineskip=12pt
  \parskip=0 pt         \parindent=20pt
  \overfullrule=0pt     \lineskip=0pt   \lineskiplimit=0pt
  \hbadness=10000 \vbadness=10000 
\newskip\ttglue
\font\fiverm=cmr5
\font\fivei=cmmi5
\font\fivesy=cmsy5
\font\fivebf=cmbx5
\font\sixrm=cmr6
\font\sixi=cmmi6
\font\sixsy=cmsy6
\font\sixbf=cmbx6
\font\ninerm=cmr9
\font\ninei=cmmi9
\font\ninesy=cmsy9
\font\nineit=cmti9
\font\ninesl=cmsl9
\font\ninett=cmtt9
\font\ninebf=cmbx9
\def\ninepoint{\def\rm{\fam0\ninerm}  
  \textfont0=\ninerm \scriptfont0=\sixrm \scriptscriptfont0=\fiverm
  \textfont1=\ninei  \scriptfont1=\sixi  \scriptscriptfont1=\fivei
  \textfont2=\ninesy  \scriptfont2=\sixsy  \scriptscriptfont2=\fivesy
  \textfont3=\tenex  \scriptfont3=\tenex  \scriptscriptfont3=\tenex
  \textfont\itfam=\nineit  \def\it{\fam\itfam\nineit}
  \textfont\slfam=\ninesl  \def\sl{\fam\slfam\ninesl}
  \textfont\ttfam=\ninett  \def\tt{\fam\ttfam\ninett}
  \textfont\bffam=\ninebf  \scriptfont\bffam=\sixbf
    \scriptscriptfont\bffam=\fivebf  \def\bf{\fam\bffam\ninebf}
  \tt  \ttglue=.5em plus.25em minus.15em
  \normalbaselineskip=11pt
  \setbox\strutbox=\hbox{\vrule height8pt depth3pt width0pt}
  \let\sc=\sevenrm  \let\big=\ninebig \normalbaselines\rm}
\pageno=0

\footline{\ifnum\pageno=0\hss\else\hss\tenrm\folio\hss\fi}
\hbox{}
\vskip 1truein\centerline{{\bf THE MOST STABLE STRUCTURE FOR HARD SPHERES}}
\vskip .2truein\centerline{by}
\vskip .2truein
\centerline{{Hans Koch
\footnote{$^1$}{Research supported in part by NSF Grant DMS-0322962},
Charles Radin
\footnote{*}{Research supported in part by NSF Grant DMS-0354994}
and Lorenzo Sadun
\footnote{**}{Research supported in part by NSF Grant DMS-0401655}}}
\vskip .2truein\centerline{Department of Mathematics, University of
  Texas, Austin, TX}

\vs2
\centerline{{\bf Abstract}}
\vs.1 \nd
The hard sphere model is known to show a liquid-solid phase
transition, with the solid expected to be either face centered cubic
or hexagonal close packed. The differences in free energy of the two
structures is very small and various attempts have been made to
determine which structure is the more stable. We contrast the
different approaches and extend one.
\vs.8
\vs.2
\vfill\eject

\vs.2 \nd {\bf I. Introduction}.  

\vs.1 

Computer simulations of the hard sphere model in the late 1950's (see
[B] for a review) showed a liquid/solid phase transition at packing
fractions in the range $0.49$ to $0.54$.  Ever since that time there
has been significant interest in determining the internal structure of
that solid phase, since the model is one of the simplest possible
showing both liquid and solid phases: the only interaction is that the
spheres may not overlap.  At least at high density, the solid is
expected to have one of the structures associated with densest packing
(packing fraction $\pi/\sqrt{18}\approx 0.74$), in particular the face
centered cubic (fcc) and hexagonal close packed (hcp) structures,
which commonly appear in materials. In 1968 Stillinger and coworkers
[RSYS] used
series expansions to try to answer the much simpler question of  
whether fcc or hcp is
more stable. They computed the first 3 terms in a series for the
entropy per sphere for both fcc and hcp and concluded that hcp had
higher entropy per sphere and was therefore more stable; no error bars
were given. We will extend this analysis to 2 more terms.
Others used molecular dynamics or Monte Carlo simulations
[HR,FL] to compare the fcc and hcp entropies, with inconclusive results.
In 1997, using more advanced computer
technology, Woodcock [W1,W2] did a molecular dynamics
simulation and concluded that fcc is more stable, with error bars to
back up this conclusion. The error bars and the magnitude of the entropy 
difference were called into question,
but further simulations [BFMH,BWA,MH] agreed with Woodcock's 
qualitative claim
that fcc is more stable than hcp. Finally, a geometric approach was
recently applied to the first nontrivial term in Stillinger's expansion [RS].
This paper concerns the effort to extend the computation of the
Stillinger expansion so as to determine the geometric mechanism 
behind stability.

\vs.2 \nd {\bf II. Analysis}

\vs.1 We consider the hard sphere model of classical statistical
mechanics in the canonical ensemble, consisting of the uniform
distribution of configurations of non-overlapping spheres at fixed
packing fraction.
(The momentum variables can be integrated out,
as they are decoupled from physical space variables in this model.) 
We assume, based on old simulations [B], that at high
density (packing fraction $d\approx \pi/\sqrt{18}\approx 0.74$)
the model is in a solid phase which is, moreover,
either an fcc or hcp crystal.
More specifically, the assumption is that
the distribution gives overwhelming weight to configurations of
spheres which are perturbations of either a perfect fcc or hcp
structure at the appropriate density. We begin with an analysis of
the meaning of ``perturbation'' and of error estimates for this
problem.

One way to make definite the meaning of perturbations of the perfect
fcc structure for $d\approx \pi/\sqrt{18}$ is as follows. Start with a
finite block of spheres frozen in a perfect fcc arrangement at density
below close packing (obtained by relative choice of sphere radius and
lattice spacing), and with a subset $Q$ of $|Q|=N$ lattice sites for which
we free up the associated spheres. As in [RSYS] we note that,
as $d\to\pi/\sqrt{18}$,
the restrictions in position of any sphere due to its
neighbors can be approximated by neglecting the curvature of the
spheres, obtaining linear restrictions. Specifically, suppose our lattice
spacing is 1, $L_j$ represents the coordinates of the $j^{th}$ lattice
site, and $u_j$ represents the relative coordinates of the center of
the sphere of diameter $1-\eps$ associated with this site. For nearest
neighbor sites $j$ and $k$, the hard
core restriction is $||L_j+u_j-L_k-u_k||\ge 1-\eps$ and the
linearization is $(L_j+u_j-L_k-u_k)\cdot (L_j-L_k)\ge 1-\eps$, or
$(u_j-u_k)\cdot(L_j-L_k)\ge -\eps$. So the entropy density $\s$ of
this class of configurations at high density is precisely $\ln(V)/N$,
where $V$ is the volume of the convex polyhedron in $\R^{3N}$ whose faces are
the linear constraints on the positions of the $N$ moveable spheres
due to the frozen spheres and each other.

Now that we have a class of configurations associated with each of fcc
and hcp at high density, we consider the corresponding entropy
densities $\s_{fcc}$ and $\s_{hcp}$. As shown in [RSYS], $\s_j(d)$
diverges as $d\to \pi/\sqrt{18}$ and is of the form
$\ln(\Delta d)+C_j+O(\Delta d)$,
where $\Delta d=\pi/\sqrt{18}-d$ and the $C_j$ are constants.
In order to determine whether fcc or hcp is more stable
one must estimate the difference $\Delta C=C_{hcp}-C_{fcc}$.
All works we are considering claim that $|\Delta C|\approx 10^{-3}$.
Using $V$ for phase space volume we have by definition
$V_j=e^{s_j N}$, so obtaining the desired accuracy in $\s_j$ to order
$10^{-3}$ for each case requires a bound on the relative error for
the volumes of $\Delta V_j/V_j<e^{N/10^3}-1$. 

A different way to make sense of perturbations of the fcc
and hcp structures is to consider deviations of these structures
that are periodic with some fixed period;
if the density is high enough the two sets of configurations
are disconnected and thus the volumes
$V_{fcc}$ and $V_{hcp}$ are well defined.
Both molecular dynamics [W1,W2] and Monte Carlo simulations
[BFMH,BWA,MH] used this setup, although mostly at densities
near melting, or equivalently, for spheres of radius about $0.9$,
where the two sets of configurations are actually connected to
one another
(for periods $6$ or larger in each direction).
At these densities, a simulation should theoretically
sweep the {\it entire} phase space, regardless of where it is started.
However, on the time scale available in practice
the process is found to be trapped in regions
whose configurations can be associated with a single lattice structure,
and it is the volumes of these regions that have been used
to conclude that fcc is the more stable structure.
Although the (extremely large) fcc-hcp mixing time may not be a problem 
in these simulations,
the success of the method does depend on the absence of 
relevant time scales beyond the ones that have been observed so far.
Moreover, such simulations
do not give any intuition into the mechanism, almost certainly
geometric, which distinguishes these close packed structures.

We now consider the method of
Stillinger et al, which does not (inherently) rely on
simulation and does in principle allow for a geometric interpretation.
The first nontrivial term in the expansion below has been analyzed in [RS].
Consider again our finite
block of spheres frozen in a perfect crystal arrangement, fcc or hcp,
at density below close packing, and a subset $Q$ of them which we free up.
The entropy $S_Q$ of the moveable spheres associated with (or labelled by)
the sites $Q$ can be written as an exact sum
$$
\S_Q= \sum_{P\subseteq Q} C_P=\sum_{P\subset
Q: |P|=1} C_P\ \ + \sum_{P\subset Q: |P|=2}C_P\ \ +\ \ \cdots
$$
which defines the correction terms $C_P$. For each $n>0$ there are,
up to symmetries, a fixed number of contributing
``polymers'' $P$ of $|P|=n$ spheres;
one can compute these and then each ``$n$-body correction''
$s_{n,Q}={1\over|Q|}\sum_{P\subseteq Q: |P|=n}C_P$, so that
$$
\s_Q={\S_Q\over |Q|}=\sum_{n\ge 1}\s_{n,Q}.
$$
As $|Q|\to\infty$ the number of occurences per site of the polymer $P$
approaches a fixed frequency $f_P$, and $\s_{n,Q}$
approaches a limit $\s_n = \sum_{|P|=n}f_P\, C_P$. 
The entropy per sphere of the hard sphere solid is then
$\s=\sum _{n\ge 1}\s_n$.

We specialize to densities very close to the maximum density, that is,
phase space volumes (and hence entropies) are computed only
to leading order in the difference $\eps$
between the lattice spacing $1$ and the sphere diameter $1-\eps$.
As noted above $\exp(\S_P)$ is the volume of a polyhedron in $3|P|$
dimensions, and the dimensions of this polyhedron all scale as $\eps$.  
In particular for
$|P|=1$,\ $\s_1=C_P$ is simply the logarithm of the scaled volume of
a Voronoi cell for a sphere,
and this value is the same for fcc as for hcp.
Note that for $|P|>1$ the (divergent) $\,\ln(\eps)\,$ terms in $C_P$ cancel, 
so $s_n$ is well-defined in the $\eps\to 0$ limit for $n>1$.

It is natural to simply truncate the
sum $\sum_n s_n$ and compute each $\s_n$ by computing the appropriate volumes.
See Table 1, below. 
The success of this method depends on the rapid decay of the
individual terms $\s_n$ with increasing $n$, so that the sum of
uncomputed terms are convincingly negligible. Note that for both fcc
and hcp, and for all computed levels, $\s_{n+1}$ is in absolute value
considerably smaller than $\s_n$.  Recalling that we want
to estimate $\Delta \s$ to within an error of $10^{-3}$, and assuming
the two series remain geometrically decreasing by a factor of roughly $3$,
one needs to compute all terms up to level $n=7$.
Unfortunately this seems unattainable with current
computers as we discuss in the next section.

\vs.1

{\ninepoint
$$
\vbox{\tabskip=0pt\offinterlineskip\halign to 303.9pt{
\strut#&\vrule#\tabskip=0.5em&\hfil#&\vrule#&\hfil#&\vrule#&\hfil#&\vrule#&\hfil#
&\vrule#&\hfil#&\vrule#&\hfil#&\vrule#\tabskip=0pt\cr
\noalign{\hrule}
&&$N$\hfil&
  &fcc:\ \ $s_n$\hfil&&fcc:\ $\sum_1^n s_m$\hfil&
  &hcp:\ \ $s_n$\hfil&&hcp:\ $\sum_1^n s_m$\hfil&
  &${\phantom{\Bigm|}}\Delta s$\hfil&\cr
\noalign{\hrule}
&&$1$\hfil&&$ 1.0986122$&&$1.0986122$&&$ 1.0986122$&&$1.0986122$&&$ 0.0000000$&\cr
\noalign{\hrule}
&&$2$\hfil&&$-0.1647410$&&$0.9338712$&&$-0.1636474$&&$0.9349648$&&$-0.0010935$&\cr
\noalign{\hrule}
&&$3$\hfil&&$-0.0517903$&&$0.8820808$&&$-0.0518249$&&$0.8831398$&&$-0.0010589$&\cr
\noalign{\hrule}
&&$4$\hfil&&$-0.0274325$&&$0.8546482$&&$-0.0261820$&&$0.8569577$&&$-0.0023094$&\cr
\noalign{\hrule}
&&$5$\hfil&&$ 0.0060605$&&$0.8607088$&&$ 0.0026023$&&$0.8595601$&&$ 0.0011486$&\cr
\noalign{\hrule}}}
$$
\centerline{{\rm Table 1. Partial entropies for fcc and hcp structures.}}
}

In addition to computing and adding terms in order, we can examine the
contributions of individual polymers.
Figure 1, below, shows the distribution of entropy corrections $C_p$
that contribute to $s_5$,
after convolution by a Gaussian with variance $10^{-10}$.
Note that the contribution of individual polymers should not be viewed as
independent random variables.  The sum of the
$\sim{}$500 terms contributing to $s_5$ for hcp is not of order
$\sqrt{500}$ larger than a typical term: rather, it is the same size
as a typical term.  Likewise for $s_3$, $s_4$, and for fcc.  Hopefully,
by studying the distribution of terms contributing to $s_3$, $s_4$ and
$s_5$ we will be able to devise resummation schemes with accelerated
convergence.

\gdef\myepsfig#1;#2;{\everypar={}\epsfxsize=#1\nd \hbox{\hskip-1in\epsfbox{#2}}}

\myepsfig.7\hsize; 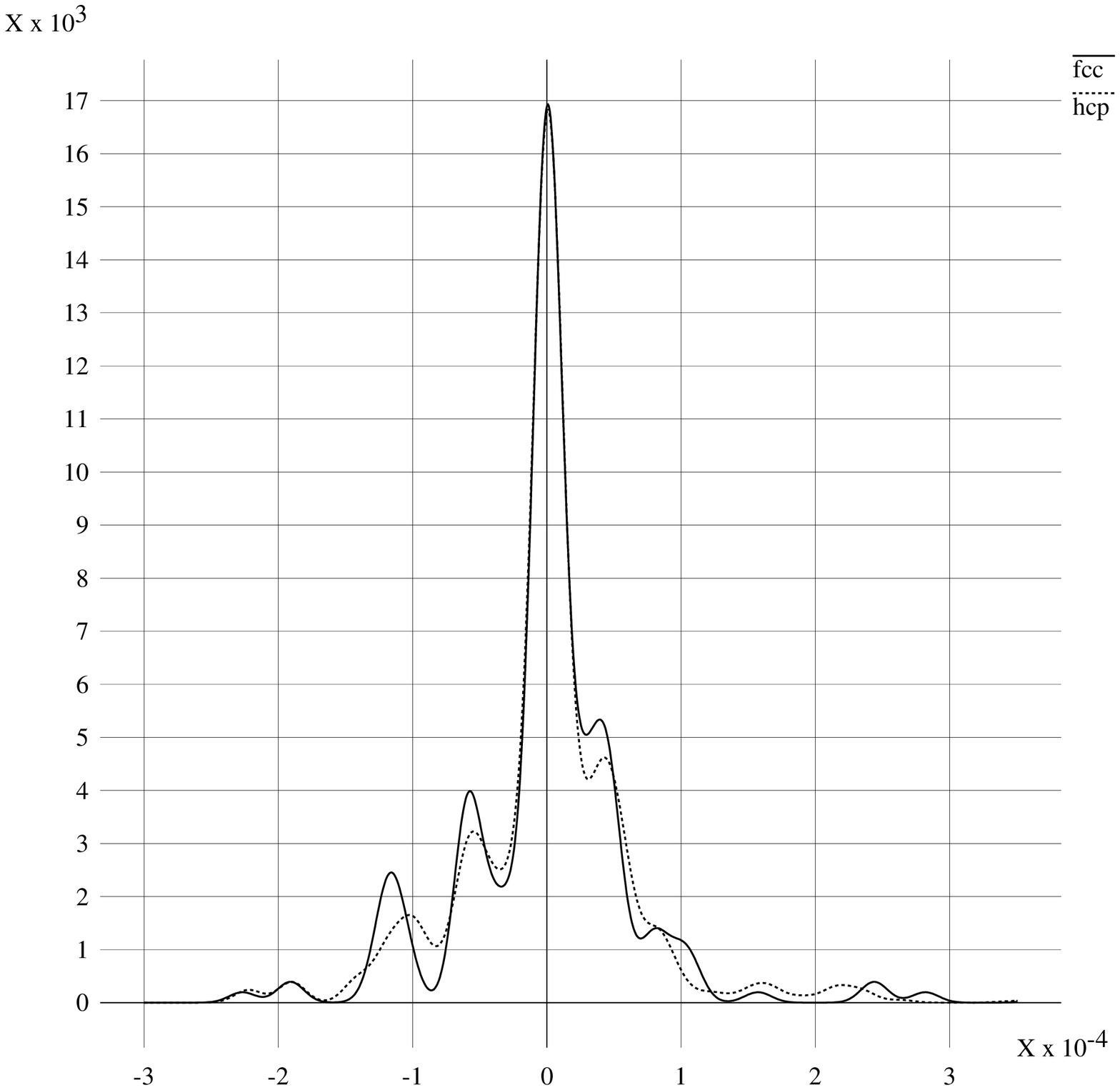;

\vs-1
{\ninepoint
\centerline{{\rm Figure 1. Distribution of entropy corrections $C_P$
contributing to $s_5$.}}
}

\vs.2

Another important goal is to understand the geometric mechanisms
underlying the difference in entropy between fcc and hcp.  The
qualitative differences between the terms contributing to $s_2$ were
considered in [RS].  Although some general patterns have emerged from
the data presented here (e.g., compact polymers tend to contribute
negatively, while extended polymers tend to contribute positively), a
true geometric understanding still eludes us.

\vs.2 \nd
{\bf III. Computation}
\vs.1 
The computations by Stillinger et al of ($\s_2$ and) $\s_3$ to 5-digit
accuracy was remarkable for the time. We have taken advantage of
current computers to compute $\s_n$ up to $n=5$. To obtain $\s_5$ we
needed to compute among other things the volumes of 591 polyhedra in
15 dimensions (the phase space for 5 spheres).  The basic algorithm is
simple, based on successive ``triangularizations'': starting with
$d=15$, choose a point in the $d$-dimensional polyhedron to serve as
the common apex of pyramids with the faces of the polyhedron as bases,
and then add up the pyramid volumes $AB/d$, where $A$ is the altitude
and $B$ the volume of the base.  Given that the base is again a
polyhedron, now in dimension $d-1$, the process can be iterated.  The
main drawback of this algorithm is that the same lower dimensional
volumes are computed repeatedly in different branches of the
recursion.  In the case of our 15-dimensional polyhedra, which have on
average 55 faces, this leads to a highly impractical number of
computations.  The standard way of dealing with this problem is to
store and to re-use the results from previous computations.  Programs
that implement this strategy are readily available, but for reasons of
efficiency (the program that was recommended to us was too slow
by several orders of magnitude)
we wrote the necessary code ourselves; it is available at xx.  Volumes in 15
dimensions each took between 7 and 160 hours of computation time, with
an average of 29 hours, on a 64-bit processor running at 2.2 GHZ,
using 2GB of memory.  The number of re-used lower dimensional volumes
in each of these cases is of the order of $10^{10}$. Graphs of the
entropy corrections $f_P\, C_P$ associated with the volumes $P$
are shown in Fig 1.

We do not expect
such an exact computation of $\s_6$ to be practical before substantial
progress in computer hardware.
\vs.2 \nd
\vs.2 \nd
\centerline{{\bf References}}
\vs.2

\item{[B]} J. A. Barker, Lattice Theories of the Liquid State (Macmillan, New
York, 1963).

\item{[BWA]} A. D. Bruce, N. B. Wilding, and G. J. Ackland,
Phys. Rev. Lett. 79, 3002 (1997).

\item{[BFMH]} P. G. Bolhuis, D. Frenkel, S.-C. Muse, and D. A. Huse, Nature
(London) 388, 235 (1997).

\item{[FL]} D. Frenkel and A. Ladd, J. Chem. Phys. 81, 3188 (1984)


\item{[HR]} W. Hoover and F. Ree, J. Chem. Phys. 49, 3609 (1968).


\item{[MH]} S.-C. Mau and D. A. Huse, Phys. Rev. E 59, 4396 (1999).


\item{[RS]} C. Radin, and L. Sadun, Phys. Rev. Lett. 94, 015502 (2005).

\item{[RSYS]} W. G. Rudd, Z.W. Salsburg, A. P. Yu, and F. H. Stillinger,
J. Chem. Phys. 49, 4857 (1968).



\item{[W1]} L.V. Woodcock, Nature (London) 385, 141 (1997).

\item{[W2]} L.V. Woodcock, Nature (London) 388, 236 (1997).

\end